\documentclass[reprint,amsmath,amssymb,aps,prstab]{revtex4-1}

\usepackage[english] {babel}
\usepackage{amsmath,amssymb,amsfonts,hyperref,graphicx,natbib,xcolor}
\usepackage{mathrsfs} 
\usepackage{amsthm}
\usepackage{feynmp}
\usepackage{dsfont}
\usepackage{siunitx}
\usepackage[all]{xy}
\usepackage[mathcal]{eucal}

\hypersetup{
  pdfauthor={Stanislav Baturin}
}

\renewcommand{\d}{\partial}

\begin{document}

\title{Analytic model of 3D beam dynamics in a wakefield device}

\author{Dae Heun Koh}%
\affiliation{The University of Chicago, PSD Enrico Fermi Institute, 5640 S Ellis Ave, Chicago, IL 60637, USA}%
\author{Stanislav S. Baturin}%
\email{s.s.baturin@gmail.com}%
\address{The University of Chicago, PSD Enrico Fermi Institute, 5640 S Ellis Ave, Chicago, IL 60637, USA}%
\date{\today}

\begin{abstract}
In this paper we suggest an analytic model,
and derive simple formulas, for the beam dynamics in a wakefield structure of arbitrary cross-section. The results could be applied to estimate an upper limit of the projected beam size in devices such as the dechirper and the wakefield streaking device. The suggested formalism is also applicable to the case when slices of the beam are distributed along an arbitrary line in 3D.     
\end{abstract}

\maketitle

\section{Introduction}

The effects of beam instabilities, or beam breakup (BBU),  resulting from parasitic wakefields, considerably limits the intensity of the beam that can be transported through accelerator structures. 
BBU simulations are a key element in the design and development of accelerator components. The most accurate and up-to-date approach to calculating BBU effects is the use of direct particle tracking, however this method often requires substantial computational resources. 
Thus, there is a demand for beam dynamics tools that provide quick estimates to crosscheck complex simulation results.
Analytic models for BBU analysis were initially developed in Refs.\cite{Panofsky,CRY,Lau}, further refined in Ref.\cite{Ilya}, and the most complete, recent solution was derived by Delayen in a series of publications Refs.\cite{Delayen1,Delayen2,Delayen3}.  

Contemporaneously, a new theoretical model that can be used for obtaining direct analytic formulas for the transverse wakefields was developed in Ref.\cite{mySTAB}. In this approach, an upper limit for the transverse and longitudinal wake forces in a beam pipe of arbitrary cross-section was derived. 
In this paper, starting from a simple form of the upper limit for the wake force derived in Ref.\cite{mySTAB}, we propose an approximate equation of motion, and derive the solution for the  transverse dynamics of a pencil-like beam (beam transverse size  much smaller then the aperture). 
In contrast to Ref.\cite{Delayen1}, we consider a simplified expression for the transverse wakefield and show that in our case, the solution is generalized to a beam distributed along a line in 3D. 
In the special case when all beam slices have the same initial displacement along the central axis, and the beam is not twisted, our results are in complete agreement with Ref.\cite{Delayen1}.  
We also provide a general expression for the instability growth length (see Ref.\cite{Schroeder} for details). As an illustrative example, we apply the developed formalism to study the beam dynamics in a cylindrical structure.   

The results  presented are general and applicable to practical accelerator components, for example, to the estimation of the upper limit of the projected beam size in devices such as the dechirper \cite{DCBane,CorPipe,DielStr} and the wakefield streaking device \cite{St1}. 

\section{Model description}

\begin{figure}[h]
\begin{center}
\includegraphics[scale=0.21]{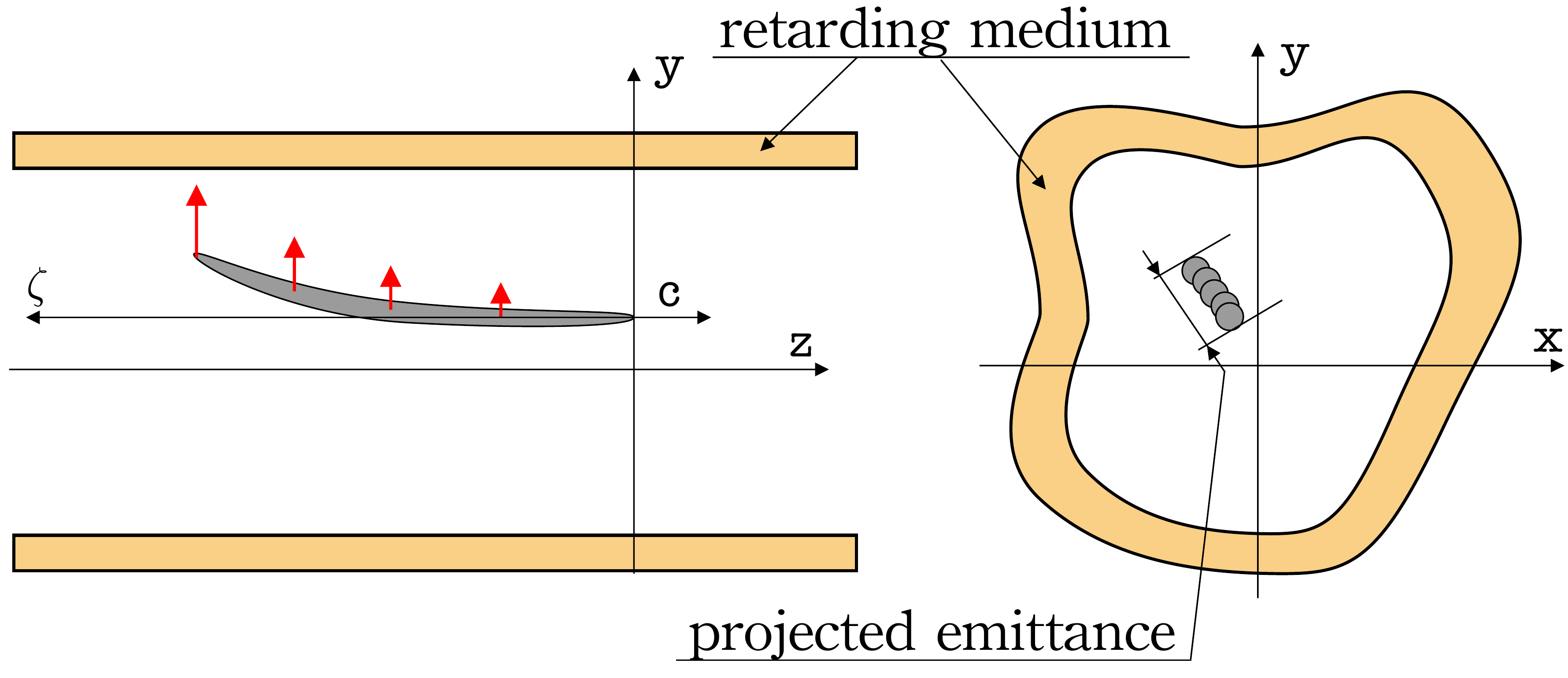}
\caption{Schematic diagram of the transverse motion of ultra relativistic particle beam.}
\label{Fig:1}
\end{center}
\end{figure}

We consider an electron beam traveling close to the speed of light along the axis of the wakefield structure (Fig.\ref{Fig:1}) with an initial displacement off the center of the structure. 

 We introduce the complex transverse velocity, momentum and force as (consistent with the notation of the Ref.\cite{mySTAB}),

\begin{align}
\label{eq:ini}
&V_\perp=V_x+iV_y, \nonumber \\
&p_\perp=p_x+ip_y ,\\
&F_\perp=F_x+iF_y. \nonumber 
\end{align}
The trasnverse equation of motion is written as

\begin{align}
\frac{\partial \gamma m_e V_\perp} {\partial t}= F_\perp,
\end{align}
where $\gamma$ is the relativistic gamma factor and $m_e$ is the electron rest mass.  
The beam is assumed to be ultra-relativistic $V_z \approx c$ and consequently $z \approx ct$, hence

\begin{align}
\frac{\partial p_\perp}{\partial t} \approx c\frac{\partial p_\perp}{\partial z}.
\end{align}
For  relativistic momentum $p_\perp = \gamma m_e V_\perp$, we have

\begin{align}
\frac{\partial \gamma}{\partial z} V_\perp + \gamma \frac{\partial V_\perp}{\partial z}=\frac{F_\perp}{m_e c}.
\end{align}
For the case when the energy loss is small compared to the device length, $d\gamma/dz \approx 0$ we have

\begin{align}
\label{eq:Neq}
\frac{\partial^2 \omega}{\partial z^2}
= \frac{F_\perp}{\gamma m_e c^2}.
\end{align}
Here $\omega=x+iy$ is the complex vector of the transverse positions. 
The transverse Lorentz force acting on the beam is usually cast in a framework of the transverse wakefield. The Lorentz force from a distributed charge is a convolution of the transverse wakefield of a point particle $G_{\perp}$, with the particle distribution,  \cite{Chao, Zotter}

\begin{align}
\label{eq:Fr}
F_\perp(\omega,\zeta) =\int G_\perp[\omega,\omega_0,\zeta,\zeta_0] \rho(x_0,y_0,\zeta_0)  dx_0dy_0d\zeta_0.
\end{align}

We consider a model of a pencil like bunch \cite{Chao} assuming that transverse size of the bunch is much smaller than the vacuum chamber aperture size. In this case, the transverse motion of different longitudinal slices of the bunch (slices by $\zeta$ coordinate, see Fig.\ref{Fig:1}) is defined by the motion of the center of mass of the slice. So we do not account for the effects of the slice shape modification, treating each slice as a point particle.  Then we approximate the particle density distribution $\rho(x,y,\zeta)$ as

\begin{align}
\label{eq:rho}
\rho(x,y,\zeta)=\delta[x(\zeta)]\delta[y(\zeta)]\rho_l(\zeta).    
\end{align}
Here $\rho_l(\zeta)$ is the longitudinal charge distribution and $\delta$  is the Dirac delta function.
Substitution of Eq.\eqref{eq:rho} into Eq.\eqref{eq:Fr} gives

\begin{align}
\label{eq:lorF}
F_\perp(\omega(\zeta),\zeta) = \int G_\perp[\omega(\zeta),\omega_0(\zeta_0),\zeta,\zeta_0]\rho_l(\zeta_0)d\zeta_0.
\end{align}
\begin{figure}[t]
\begin{center}
\includegraphics[scale=0.25]{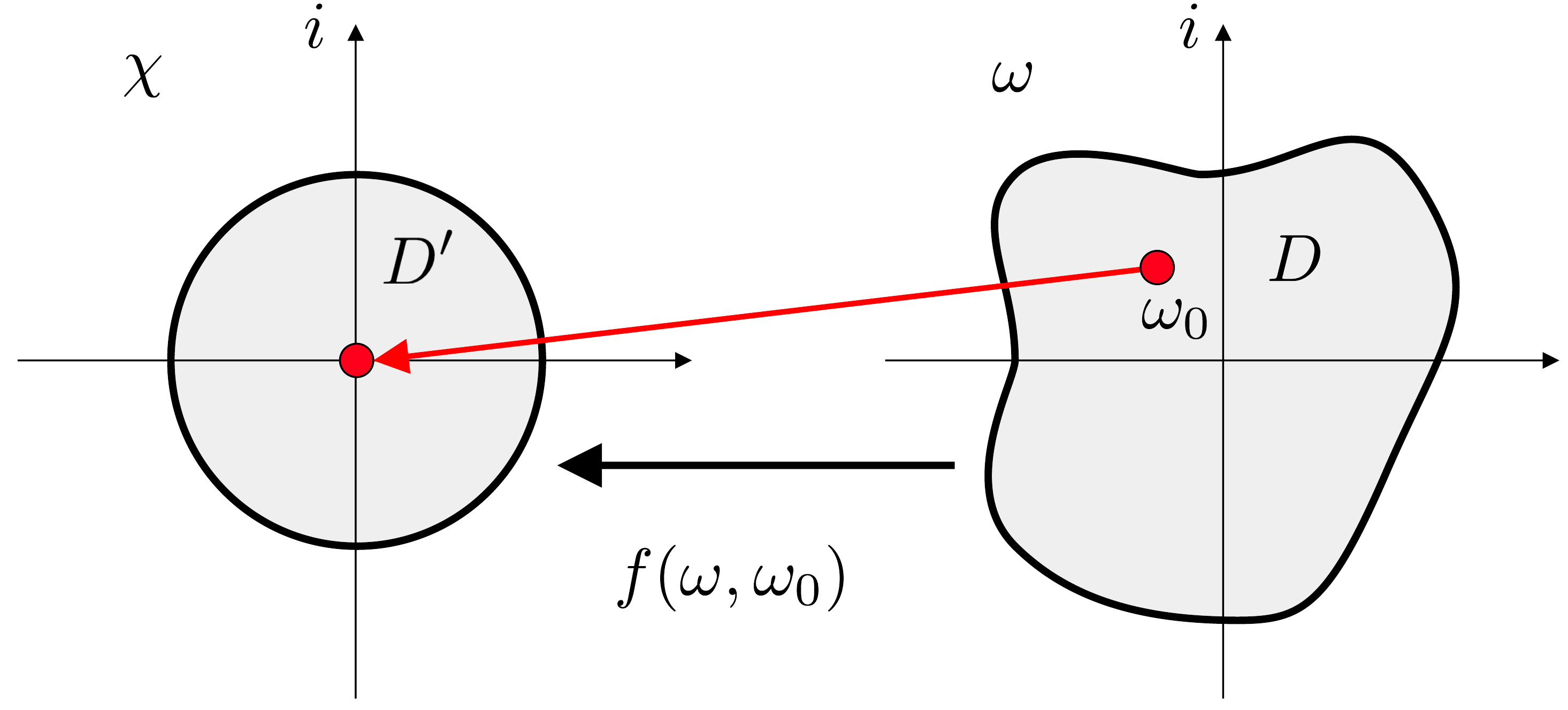}
\caption{Schematic diagram of the conformal mapping.}
\label{Fig:conf}
\end{center}
\end{figure}
It was shown in Ref.\cite{mySTAB} (see Section 3 and Appendix B of the Ref.\cite{mySTAB} for the details) that the upper limit of the Green's function $G_\perp$ for an arbitrary cross section of longitudinally homogenous vacuum channel could be written in terms of a conformal mapping function $f(\omega,\omega_0)$ that maps the cross sectional area onto a unit disk (see Fig.\ref{Fig:conf}). Here $\omega=x+iy$ is the position of the point test particle and $\omega_0=x_0+iy_0$ is the position of the point source particle. Mapping is arranged in such a way that $\omega_0$ corresponds to the center of the unit disk. The corresponding expression in complex notation could be written according to Ref.\cite{mySTAB} as

\begin{align}
\label{final_Lor0}
&G_\perp(\omega,\omega_0,\zeta,\zeta_0)\leq  \\ &\frac{4 eQ \theta(\zeta-\zeta_0)}{a^2}(\zeta-\zeta_0)f''(\omega,\omega_0)^* f'(\omega_0,\omega_0) \nonumber.
\end{align}
Here $Q$ is the total charge of the bunch, $e$ is the electron charge, and $a$ is the characteristic size of the aperture. The force is given in CGS units.

For small displacements $\delta\omega$ and $\delta\omega_0$ from the initial position $\omega_0$ and point $\omega$ we may linearize right hand side of the Eq.\eqref{final_Lor0} as

\begin{align}
\label{final_Lor1}
&G_\perp(\omega,\omega_0,\zeta,\zeta_0)\lesssim  \\ &(\zeta-\zeta_0)\theta(\zeta-\zeta_0)\left[ \tilde{A}+ \tilde{B}\delta\omega_0(\zeta_0)+\tilde{C}\delta\omega(\zeta)^*\right] \nonumber,
\end{align}  
with

\begin{align}
\label{eq:const}
\tilde{A}&=\frac{4 eQ} {a^2} f''(\omega_0,\omega_0)^* f'(\omega_0,\omega_0), \nonumber \\ 
\tilde{B}&=\frac{4 eQ} {a^2} \left.\frac{\partial f''(\omega,\omega_0)^* f'(\omega_0,\omega_0)}{\partial \omega_0}\right|_{\omega=\omega_0}, \\
\tilde{C}&=\frac{4 eQ} {a^2}\left.\frac{\partial f''(\omega,\omega_0)^* f'(\omega_0,\omega_0)}{\partial \omega^*}\right|_{\omega^*=\omega_0^*}.\nonumber
\end{align} 
Consequently for the case of $\delta\omega=\delta\omega_0=\bar\omega$ when transverse force is taken at the bunch position we have:

\begin{align}
\label{final_Lor}
&G_\perp(\zeta,\zeta_0)\lesssim(\zeta-\zeta_0)\theta(\zeta-\zeta_0)\left[ \tilde{A}+\tilde{B}\bar\omega(\zeta_0)+\tilde{C}\bar\omega(\zeta)^*\right].
\end{align}  
We the substitute Eq.\eqref{final_Lor} into Eq.\eqref{eq:Neq}, and with Eqs.\eqref{eq:lorF} we arrive at

\begin{align}
\label{eq:mainM}
&\frac{\partial^2 \bar\omega(\zeta,z)}{\partial z^2}
=[A+C \bar\omega(\zeta,z)^*]\int_0^\zeta\rho_l(\zeta_0) (\zeta-\zeta_0) d\zeta_0+ \nonumber \\ &B \int_0^\zeta \bar\omega(\zeta_0,z)\rho_l(\zeta_0) (\zeta-\zeta_0) d\zeta_0.
\end{align}  
For  convenience we introduce $A =\frac{\tilde{A}}{\gamma_0 m_e c^2}$,  $B =\frac{\tilde{B}}{\gamma_0 m_e c^2}$ and $C =\frac{\tilde{C}}{\gamma_0 m_e c^2}$. 

By comparing  Eq.\eqref{eq:mainM} with the commonly used equation of the 1D transverse motion (see for example Ref.\cite{Chao}) we observe that the introduced constants $A$,$B$,$C$ are directly related to the monopole, dipole and quadrupole wakefield amplitude respectively.  It should be noted that the monopole amplitude, $A$, may appear due to the intentionally large initial displacement from the axis of symmetry of the waveguide or due to a general fundamental asymmetry of the waveguide.         

Eq.\eqref{eq:mainM} is the central equation of the suggested model. This equation, and the coefficients in this equation, are valid for arbitrary waveguide of arbitrary material.  The complex function $\omega$ that yields the solution to this equation is essentially an upper bound for any real trajectory. The real trajectory will be bounded by the limiting trajectory as $|\omega_{rl}|\leq|\omega|$.

\section{Analytic solution to the dynamics problem}

Brief analysis of  Eq.\eqref{eq:mainM} suggests that a closed form analytic solution to  Eq.\eqref{eq:mainM} probably does not exist.  
However, for special cases when the dipole wakefield dominates, the quadrupole wakefield $|\tilde{B}|>>|\tilde{C}|$ and charge distribution is uniform $\rho_l(\zeta)=1/l$ , $\zeta\in [0,l]$ a simple closed from solution could be found. 

Under these assumptions, according to Eq.\eqref{eq:Neq}, Eq.\eqref{eq:lorF} and Eq.\eqref{final_Lor}, the equation of motion is reduced to:

\begin{align}
\label{eq:mot}
\frac{\d^2 \bar\omega(\zeta,z)}{\d z^2} =\frac{A \zeta^2}{2l}+\frac{B}{l} \int_0^\zeta \bar\omega(\zeta_0,z) (\zeta-\zeta_0) d\zeta_0.
\end{align}
with initial conditions:

\begin{align}
\label{eq:incond}
\bar\omega(\zeta,0) = l_0(\zeta), \quad \frac{\d \bar\omega(\zeta,0)}{\d z} = v_0(\zeta).
\end{align}

\begin{figure}
\begin{center}
\includegraphics[scale=0.16]{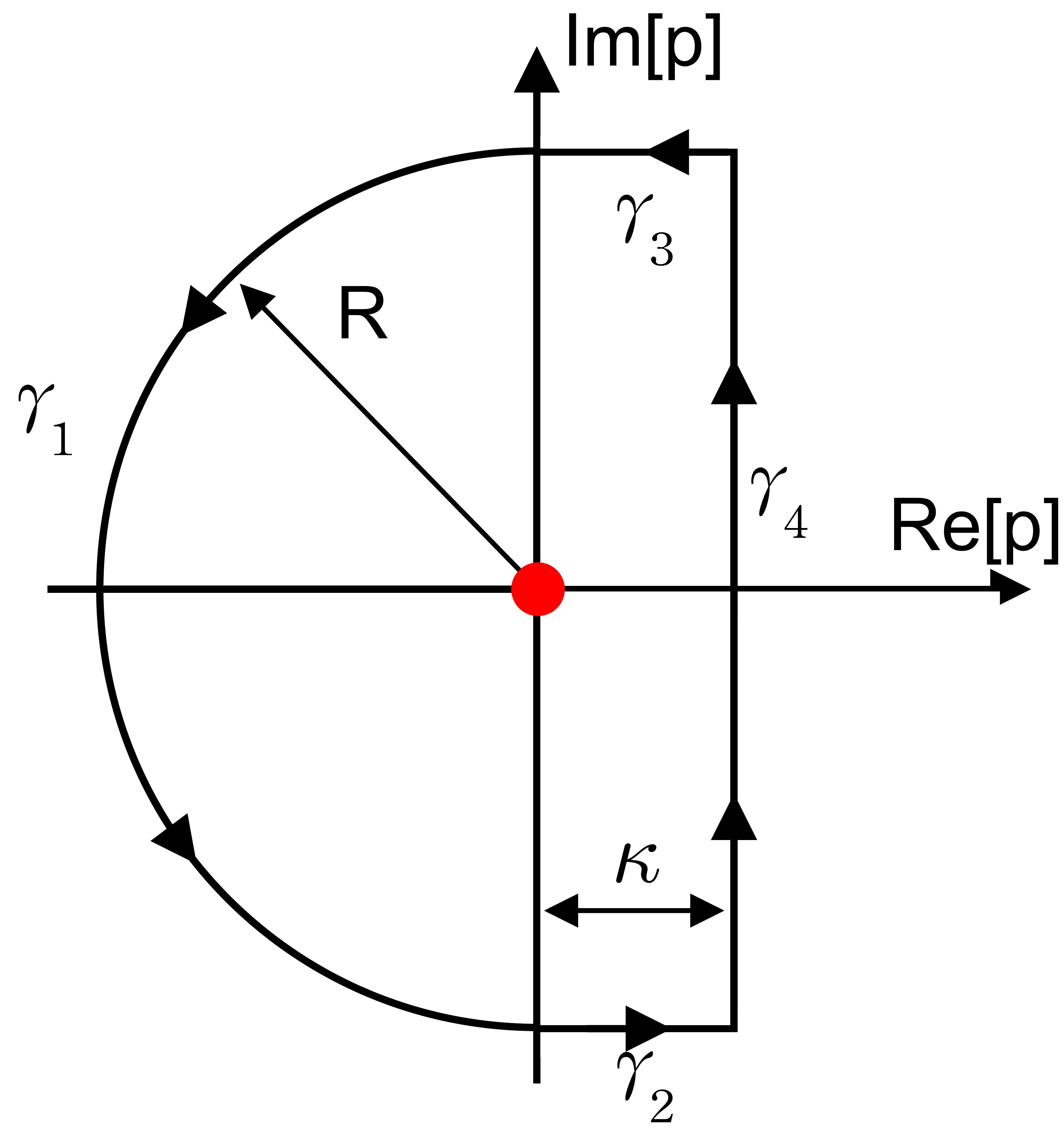}
\caption{Contour integral path for computing the inverse Laplace transform.}
\end{center}
\label{fig:2}
\end{figure}

We define forward and inverse Laplace transformations as \cite{shabat}

\begin{align}
\label{eq:lap}
\mathscr{L}[f(\zeta)]=\int\limits_{0}^\infty f(\zeta)e^{-p\zeta}d\zeta
\end{align}
and

\begin{align}
\label{eq:ilap}
\mathscr{L}^{-1}[F(p)] = \frac{1}{2 \pi i} \lim \limits_{R \to \infty} \int_{\kappa-iR}^{\kappa+iR} e^{\zeta p} \ F(p) \ dp,
\end{align}
where $\kappa$ is a real number such that the line $p = \kappa$ in the complex plane avoids singularities of $F(p)$.
We perform Laplace transformation by $\zeta$ on both sides of  Eq.\eqref{eq:mot}. On the left hand side we have:

\begin{align}
\mathscr{L}\left[ \frac{\d^2 \bar\omega(\zeta,z)}{\d z^2}\right] = \frac{\d^2}{\d z^2} \mathscr{L}\left[\bar\omega(\zeta,z)\right].
\end{align}
We define $\Omega(p,z)$ as

\begin{align}
\Omega(p,z) := \mathscr{L}\left[\bar\omega(\zeta,z)\right].
\end{align}
Hence, we have

\begin{align}
\frac{\d^2 \Omega(p,z)}{\d z^2} = \mathscr{L}\left[\frac{A\zeta^2}{2l}+\frac{B}{l}\int\limits_0^\zeta \bar\omega(\zeta_0,z)(\zeta-\zeta_0) \ d\zeta_0 \right].
\end{align}
Using the convolution-multiplication theorem for the Laplace transformation \cite{shabat}: 

\begin{align}
\mathscr{L}\left[\int\limits_{0}^\zeta g(\zeta) f(\zeta-t) dt \right]=\mathscr{L}[g(\zeta)]\mathscr{L}[f(\zeta)],
\end{align}
we rewrite the integral on the right side of Eq.\eqref{eq:mot} in a Laplace image as 

 \begin{align}
\label{eq:Lim0}
\frac{\d^2 \Omega(p,z)}{\d z^2} =\frac{A}{lp^3}+\frac{B}{l p^2} \Omega(p,z)
\end{align}
with the Laplace image of the integral kernel $\mathscr{L}[\zeta]$ given by

\begin{align}
\mathscr{L}[s]= -\frac{1}{p} \zeta e^{-p\zeta} \bigg|_0^\infty + \int_0^\infty \frac{e^{-p\zeta}}{p} d\zeta = \frac{1}{p^2}.
\end{align}
General and partial solutions to the Eq.\eqref{eq:Lim0} are

\begin{align}
&\Omega_g(p,z) = f_1(p) \ e^{\frac{\sqrt{A}z}{p}} + f_2(p) \ e^{-\frac{\sqrt{A} z}{p}}, \\
&\Omega_{pt}(p,z) =-\frac{A}{B p}, 
\end{align}
where $f_1$ and $f_2$ are functions of $p$. We define the Laplace image for the initial conditions \eqref{eq:incond} as:

\begin{align}
\label{eq:fun}
L_0(p) = \mathscr{L}\left[l_0(\zeta)\right]
\end{align}
and

\begin{align}
\label{eq:deriv}
V_0(p) = \mathscr{L}\left[v_0(\zeta)\right].
\end{align}
Therefore, the solution in the Laplace image is given by

\begin{align}
\label{eq:Lim}
\Omega(p,z)& =-\frac{A}{B p}+\left(\frac{A}{B}+pL_0(p)\right)\frac{\cosh{\left(\frac{\sqrt{B}z}{\sqrt{l}p} \right)}}{p} \nonumber\\&+\frac{pV_0(p)\sqrt{l}}{\sqrt{B}}\sinh{\left(\frac{\sqrt{B}z}{\sqrt{l}p} \right)}.
\end{align}
Now, we introduce the following notation to isolate the functions,

\begin{align}
 g_l(\zeta,z)=\mathscr{L}^{-1}\left[\frac{\cosh{\left(\frac{\sqrt{B}z}{\sqrt{l}p} \right)}}{p}\right]
 \end{align}
and  
  
  \begin{align}
  g_v(\zeta,z)=\mathscr{L}^{-1}\left[\sinh{\left(\frac{\sqrt{B}z}{\sqrt{l}p} \right)}\right].
  \end{align} 
Next, we focus on computing the functions $g_{l}(\zeta,z)$ and $g_{v}(\zeta,z)$. 
 We note that $\frac{1}{p}\cosh{\left(\frac{\sqrt{B}z}{\sqrt{l}p} \right)}$  has an essential singularity at $p = 0$. We use the fact that $\cosh(x)$ is an order one entire function.
As a consequence it could be represented by its Taylor series everywhere, including in small vicinity of the point $x=\infty$, thus we use the identity
 
 \begin{align}
\label{eq:Lim20}
\frac{\cosh{\left(\frac{\sqrt{B}z}{\sqrt{l}p} \right)}}{p} = \frac{1}{p} \sum\limits_{n=0}^{\infty}\frac{1}{(2n)!}\left[\frac{\sqrt{B/l}z}{p}\right]^{2n}.
\end{align}    
Using the definition of the inverse Laplace transformation Eq.\eqref{eq:ilap} we write for $g_l(\zeta,z)$:

 \begin{align}
\label{eq:Lim21}
&g_l(\zeta,z)= \\ \nonumber&\frac{1}{2 \pi i} \lim \limits_{R \to \infty} \int\limits_{\kappa-iR}^{\kappa+iR} e^{\zeta p}\sum\limits_{n=0}^{\infty}\frac{1}{(2n)!}\left[\frac{\sqrt{B/l}z}{p}\right]^{2n} \frac{dp}{p}.
\end{align} 
As far as the Taylor series converges uniformly, we can interchange the sum and integral and arrive at

 \begin{align}
\label{eq:Lim2}
&g_l(\zeta,z)= \\ \nonumber&\frac{1}{2 \pi i}\sum\limits_{n=0}^{\infty}\frac{1}{(2n)!} \lim \limits_{R \to \infty} \int\limits_{\kappa-iR}^{\kappa+iR} e^{\zeta p}\left[\frac{\sqrt{B/l}z}{p}\right]^{2n} \frac{dp}{p}.
\end{align}   
Now, we focus on the evaluation of the integral

\begin{align}
I_1=  \lim \limits_{R \to \infty}\int\limits_{\kappa-iR}^{\kappa+iR} e^{\zeta p}\left[\frac{\sqrt{B/l}z}{p}\right]^{2n} \frac{dp}{p}.
\end{align}
We chose the integration path as shown on Fig.\ref{fig:2}. According to  Jordan's lemma \cite{shabat}, we write the 
$\gamma_1$ part of the contour as

\begin{align}
\label{eq:g1}
\left|\int\limits_{\gamma_1} e^{\zeta p}\left[\frac{\sqrt{B/l}z}{p}\right]^{2n} \frac{dp}{p}\right|\leq \frac{|B|^n\pi}{l^n}\frac{z^{2n}}{R^{2n}},
\end{align}
and for the integral along the $\gamma_2$ and $\gamma_3$ paths
 
\begin{align}
\label{eq:g23}
\left|\int\limits_{\gamma_{2,3}}e^{\zeta p}\left[\frac{\sqrt{B/l}z}{p}\right]^{2n}\frac{dp}{p}\right|\leq \kappa  e^{\kappa \zeta}\frac{|B|^n z^{2n}}{l^nR^{2n+1}}.
\end{align}
From Eq.\eqref{eq:g1} and Eq.\eqref{eq:g23} in the limit  of $R\to \infty$ we see that paths $\gamma_1$, $\gamma_2$  and $\gamma_3$ do not contribute to the integral. With this one may write

\begin{align}
I_1=\oint\limits_{\gamma} e^{\zeta p}\left[\frac{\sqrt{B/l}z}{p}\right]^{2n} \frac{dp}{p}.
\end{align}
with $\gamma=\gamma_1\cup\gamma_2\cup\gamma_3\cup\gamma_4$.

We notice that integrand has a pole of the $2n+1$ order at the point $p=0$.
Hence, by the residue theorem (see for example \cite{shabat,silverman}), we have

\begin{align}
\label{eq:intres}
I_1=2\pi i\frac{(B/l)^n z^{2n}}{(2n)!}\lim\limits_{p\to0}\frac{d^{2n}e^{\zeta p}}{dp^{2n}}.
\end{align}
We combine Eq.\eqref{eq:Lim2} with  Eq.\eqref{eq:intres} and arrive at
\begin{align}
\label{eq:sols}
g_l(\zeta,z)=\sum\limits_{n=0}^{\infty} \frac{(B/l)^n (\zeta z)^{2n}}{(2n)!(2n)!}.
\end{align}

Following the same steps as above one may arrive at an equation for $g_v(\zeta,z)$ in the form

\begin{align}
\label{eq:sols2}
g_v(\zeta,z)=\sum\limits_{n=0}^{\infty} \frac{(B/l)^{n+1/2}z(\zeta z)^{2n}}{(2n+1)!(2n)!}.
\end{align}
The series expansions presented above could be expressed in terms of modified Bessel functions $I_0(x)$, $I_1(x)$ and Bessel functions $J_0(x)$, $J_1(x)$.

Considering a Taylor series (see for example \cite{shabat}) for $I_0(x)$, $I_1(x)$  

\begin{align}
\label{eq:I0}
&I_0(x) = \sum_{n = 0}^\infty \frac{(x /2)^{2n}}{(n!)^2}, \\
&I_1(x) = \sum_{n = 0}^\infty \frac{(x /2)^{2n+1}}{n!(n+1)!}, \nonumber
\end{align}
and for $J_0(x)$, $J_1(x)$ 

\begin{align}
\label{eq:J0}
&J_0(x) = \sum_{n = 0}^\infty (-1)^n\frac{(x /2)^{2n}}{(n!)^2}, \\
&J_1(x) = \sum_{n = 0}^\infty (-1)^n\frac{(x /2)^{2n+1}}{n!(n+1)!}. \nonumber
\end{align}
As far as these series have infinite radius of convergence, the series representation is exact for any $x$.
Substituting $x = 2(B/l)^{1/4} \sqrt{\zeta z}$,
and combining Eq.\eqref{eq:I0} and Eq.\eqref{eq:J0} 
while taking into account Eq.\eqref{eq:sols} and Eq.\eqref{eq:sols2}, we finally arrive at

\begin{align}
\label{eq:dynkerl}
&g_l(\zeta,z)=\frac{1}{2} \left[I_0(2\sqrt[4]{B/l}\sqrt{\zeta z}) + J_0(2\sqrt[4]{B/l}\sqrt{\zeta z}) \right], \\
&g_v(\zeta,z)=\frac{\sqrt[4]{B/l}\sqrt{z}}{2\sqrt{\zeta}} \left[I_1(2\sqrt[4]{B/l}\sqrt{\zeta z}) + J_1(2\sqrt[4]{B/l}\sqrt{\zeta z}) \right]. \nonumber
\end{align}
We notice that 

\begin{align}
\label{eq:not}
\mathscr{L}^{-1}[p\mathscr{L}[f(\zeta)]]=f'(\zeta)+f(0),
\end{align}
where prime denotes total derivative by $\zeta$.

Applying an inverse Laplace transformation to Eq.\eqref{eq:Lim} and using the first multiplication theorem \cite{shabat} for  Laplace transformations with Eq.\eqref{eq:not},  Eq.\eqref{eq:fun} and Eq.\eqref{eq:deriv} we arrive at

\begin{align}
\label{eq:omf}
&\bar\omega(\zeta,z) =-\frac{A}{B}+\left(l_0(0)+\frac{A}{B}\right)g_l(\zeta,z)+v_0(0)g_v(\zeta,z)+ \nonumber\\&\int\limits_0^\zeta l'_0(\zeta-\zeta_0)g_l(\zeta_0,z)d\zeta_0+ \frac{\int\limits_0^\zeta v'_0(\zeta-\zeta_0)g_v(\zeta_0,z)d\zeta_0}{\sqrt{B/l}}. 
\end{align}
Equation \eqref{eq:omf}  gives the solution to Eq.\eqref{eq:mot} with initial conditions of Eq.\eqref{eq:incond}. Eq.\eqref{eq:omf} gives a solution to a 2D problem. Motion along each individual coordinate could be derived from Eq.\eqref{eq:omf} by taking the real or imaginary part in Eq.\eqref{eq:omf}: $x(\zeta,z)=\Re [\omega(\zeta,z)]$ and $y(\zeta,z)=\Im[\omega(\zeta,z)]$. For the 1D case (setting constant $B$ to be a pure real number and the constant $A=0$), and assuming that the initial conditions do not depend on $\zeta$, the formula above gives exactly the same result as that derived earlier by Delayen in \cite{Delayen1} using different methods.      

\section{Conclusion}

We have rigorously derived an analytic formula for the evolution of a bunch that is distributed along a line in 3D. The developed formalism and model in this paper are complimentary to the previous results of Ref.\cite{Delayen1}. We have arrived at the same conclusion using a different method and demonstrated that this formula is extendable to the 3D case.  

The final result of Eq.~\eqref{eq:omf} is useful to estimate the upper limit of the bunch deformation, and applicable to structures of arbitrary cross-section. As a consequence of the theory developed in Ref.\cite{mySTAB}, this formula is also valid for arbitrary material in a slow-wave structure (dielectric, corrugation, linear plasma).  

This study serves as a first step to a more general analysis that could be performed by detailed examination of Eq.\eqref{eq:mainM}. In the current study, we neglected the "quadruple" component of the transverse force ($C=0$ in Eq.\eqref{eq:mainM}). The natural next step is the consideration of the omitted term in the equation of transverse motion, to analyze different possibilities of  instability suppression--for example, by  exploring possible beam self-damping using the BNS condition \cite{Chao}.       

\begin{acknowledgments}
This work was supported by the U.S. National Science Foundation under Award No. PHY-1549132, the Center for Bright Beams and under Award No. PHY-1535639. We would like to thank Gerard Andonian for his help with the manuscript.  
\end{acknowledgments}

\bibliographystyle{ieeetr}
\bibliography{Dynamic_eq}

\end{document}